# An advanced heat transfer model for Eulerian–Lagrangian simulations of industrial gas–solid flow systems

Toshiki Imatani[1*], Mikio Sakai[1**]

*1 Department of Nuclear engineering and management, School of Engineering, The University of Tokyo, 7-3-1 Hongo, Bunkyo-ku, Tokyo 113-8656, Japan*

\* Corresponding authors:

\* toshiki_imatani@dem.t.u-tokyo.ac.jp (Toshiki Imatani)

\*\* mikio_sakai@n.t.u-tokyo.ac.jp (Mikio Sakai)

**Abstract**

The discrete element method (DEM) coupled with computational fluid dynamics (CFD), has been developed to simulate complex solid–fluid flow systems. Today, DEM is regarded as an established approach, with extensive applications in industrial systems. Heat transfer modeling might be essential to the DEM as the industrial applications. However, existing DEM heat transfer models have fundamental limitations. These issues arise from the soft spring model inherent in DEM, where heat conduction is mathematically influenced by the spring constant. Consequently, complex modeling, considering contact state such as contact area and duration, is typically required to estimate heat conduction accurately. Moreover, the current heat transfer models exhibit poor compatibility with scaling laws, such as the coarse-grained DEM, leading to amplified temperature errors relative to motion errors. To address these challenges, we develop a novel heat transfer model based on an Eulerian framework within DEM simulations. In our approach, the Eulerian description is applied to the heat transfer calculation, while particle motion remains treated by the DEM. Notably, the heat conduction in the solid phase is captured through a simple setup by specifying the void fraction, rather than relying on complex contact modeling. The adequacy of the proposed model is demonstrated through validation tests in gas–solid flow systems, showing that the temperature distribution is independent of the particle contact state. Furthermore, the model exhibits strong compatibility with coarse-grained DEM, maintaining accuracy even at reduced computational costs. These results establish the new model's reliability and universality, positioning it as a promising standard for DEM–CFD simulations in industrial applications.

1. Introduction

Solid-fluid mixture systems are ubiquitous in industries, in particular, solid-fluid multiphase flows involving heat transfer are encountered and have been extensively studied [1–5]. Numerical simulations are frequently applied to better understand these solid-fluid mixture systems. In numerical simulations, the discrete element method (DEM) [6] coupled with computational fluid dynamics (CFD) [7] has been employed to simulate macroscopic characteristics of the solid-fluid mixture systems. Currently, the DEM is regarded as an established approach for the modeling and simulation of granular flows, and its industrial applications [8–11] have been extensively studied.

While the DEM can simulate the complex behavior of the solid particles, its applicability to heat transfer problems is limited. This limitation arises because heat conduction is mathematically influenced by the contact state of the computational particles because a soft spring model [12–18] is typically employed in the DEM. Specifically, the soft spring model might overestimate the duration of the particle contact as well as the contact area. Hence, complex heat transfer models have been developed to estimate heat conduction in the DEM, modeling the contact area [19] as well as both the actual contact area and contact duration [20,21] to correspond to the soft spring model. In addition, when the coarse-grained DEM [22–27] is employed, the error in heat transfer within the solid phase tends to be larger than that due to the particle motion, because the error in heat transfer is estimated on the error in the particle motion. This means the existing heat transfer model for the DEM has low compatibility with the coarse-grained DEM. Thus, the existing heat transfer models have the essential problems regarding heat conduction modeling and its integration into the coarse-grained model. This implies that developing a reliable heat transfer model for the DEM is challenging.

To resolve these inherent problems in heat transfer modeling for the DEM, a novel heat transfer model is developed, where the heat transfer for the solid phase is modeled by an Eulerian framework in the DEM simulation. In the new heat transfer model, the temperature of the solid phase is calculated by not using Lagrangian approach with complex modeling but using the Eulerian framework. By employing the Eulerian framework in the heat transfer for the DEM simulation, we can perform the heat transfer simulation for solid phase with simple setup, namely by setting only the void fraction without giving the contact state of the solid particles. The adequacy of the new heat transfer model is demonstrated through several validation tests in gas-solid flow systems, showing that the temperature distribution is not influenced by the soft spring value for the DEM and demonstrating good agreement between the computational and experimental results. Furthermore, the compatibility of the new heat transfer model with the coarse-grained DEM is also shown through the validation tests.

Thus, the new heat transfer model can compute the temperature distribution of the solid phase solely by estimating the void fraction, without modeling the contact state. Consequently, the reliability and universality of the proposed heat transfer model are demonstrated, positioning it as a promising approach for the DEM simulations owning to its high accuracy, broad applicability, and simple setup.

2. Numerical model

In the present study, the new heat transfer model is incorporated into our in-house code, FELMI. FELMI is originally equipped with a DEM–CFD solver and a scalar-field-based wall boundary model that combines the Signed Distance Function (SDF) with the Immersed Boundary Method (IBM). This chapter describes not only an overview of FELMI but also the details of the new heat transfer model.

2.1. DEM–CFD method

### 2.1.1. Solid phase

In this study, the Discrete Element Method (DEM) [6], incorporating the coarse-grained model [22], is applied to simulate the motion of solid particles. Adequacy of the coarse-grained DEM has been demonstrated through various validation studies [26,28–33]. In the coarse-grained DEM, one modeled particle represents $l^3$ original particles when the coarse-grained particle is $l$ times larger than the original particle. Note that $l$ is referred to as the coarse grain ratio. The equations for the translational and rotational motion of a coarse-grained particle are given as:

$$m_p^{CGM} \frac{d\boldsymbol{v}^{CGM}}{dt} = \sum \boldsymbol{F}_C^{CGM} + \boldsymbol{F}_d^{CGM} - V_p^{CGM} \nabla p + m_p^{CGM} \boldsymbol{g}, \tag{1}$$

and

$$I_p^{CGM} \frac{d\boldsymbol{\omega}^{CGM}}{dt} = \sum \boldsymbol{T}_C^{CGM}, \tag{2}$$

where the superscript $CGM$ represents the coarse-grained particles, and $m_p$, $I_p$, $V_p$, $\boldsymbol{v}$, $\boldsymbol{\omega}$, $\boldsymbol{F}_C$, $\boldsymbol{F}_d$, $\boldsymbol{T}_C$, $p$, and $\boldsymbol{g}$ are the mass of the particle, moment of inertia, volume, velocity, angular velocity, contact force, fluid drag force, torque due to contact force, fluid pressure, and gravitational acceleration, respectively. In the coarse-grained DEM, the following relationships can be given because of the assumption of the total energy agreement between the coarse-grained particles and the original particles.

$$d_p^{CGM} = l d_p^O, \tag{3}$$

$$V_p^{CGM} = l^3 V_p^O, \tag{4}$$

$$m_p^{CGM} = l^3 m_p^O, \tag{5}$$

and

$$I_p^{CGM} = l^5 I_p^O, \tag{6}$$

where the superscript $O$ represents the original particles, and $d_p$ is the diameter of the particle. The velocity and angular velocity of the coarse-grained particle are modeled as:

$$v^{CGM} = \bar{v}^O, \tag{7}$$

and

$$\omega^{CGM} = \frac{\bar{\omega}^O}{l}, \tag{8}$$

where upper bar indicates average. Hence, the contact force acting on the coarse-grained particle is given as:

$$F_C^{CGM} = F_{C_n}^{CGM} + F_{C_t}^{CGM}, \tag{9}$$

$$F_{C_n}^{CGM} = -l^3 k \delta_n^{CGM} - l^3 \eta v_n^{CGM}, \tag{10}$$

and

$$F_{C_t}^{CGM} = \begin{cases} -l^3 k \delta_t^{CGM} - l^3 \eta v_t^{CGM} & |F_{C_t}^{CGM}| \leq \mu |F_{C_n}^{CGM}| \\ -\mu |F_{C_n}^{CGM}| \dfrac{v_t^{CGM}}{|v_t^{CGM}|} & |F_{C_t}^{CGM}| > \mu |F_{C_n}^{CGM}| \end{cases}, \tag{11}$$

where $F_{C_n}$, $F_{C_t}$, $\delta_n$, $v_n$, $\delta_t$, $v_t$, $k$, $\eta$, and $\mu$ are the normal direction contact force, tangential direction contact force, normal direction overlap, normal direction relative velocity, tangential direction overlap, tangential direction relative velocity, spring constant, viscous damping coefficient, and friction coefficient, respectively. When the fluid drag force acting on the coarse-grained particle is calculated, Wen and Yu's equation ($\varepsilon > 0.8$) [34] and Ergun's equation ($\varepsilon \leq 0.8$) [35] is applied as:

$$F_d^{CGM} = \frac{\beta V_p^{CGM}}{1 - \varepsilon}(u - v^{CGM}), \tag{12}$$

$$\beta = \begin{cases} 150 \dfrac{(1-\varepsilon)^2}{\varepsilon} \dfrac{\mu_f}{d_p^{O^2}} + 1.75(1-\varepsilon) \dfrac{\rho_f}{d_p^O} |u - v^{CGM}| & (\varepsilon \leq 0.8) \\ 0.75 C_d \dfrac{\varepsilon(1-\varepsilon)}{d_p^O} \rho_f |u - v^{CGM}| \varepsilon^{-2.65} & (\varepsilon > 0.8) \end{cases}, \tag{13}$$

$$C_d = \begin{cases} \dfrac{24}{\text{Re}_p}(1 + 0.15 \text{Re}_p^{0.687}) & (\text{Re}_p \leq 1000) \\ 0.44 & (\text{Re}_p > 1000) \end{cases}, \tag{14}$$

and

$$\text{Re}_p = \frac{\rho_f d_p^O |\boldsymbol{u} - \boldsymbol{v}^{CGM}|}{\mu_f}, \tag{15}$$

where $\varepsilon$, $\beta$, $C_d$, $\text{Re}_p$, $\rho_f$, $\mu_f$, $\boldsymbol{u}$, and $d_p$, are the void fraction, momentum exchange coefficient, drag coefficient, Reynolds number for the particle, fluid density, viscosity, velocity, and particle diameter, respectively.

2.1.2. Fluid phase

The governing equations of the fluid are the Navier-Stokes equations and the continuity equation, where local volume averaging [36] is employed, are given as:

$$\frac{\partial(\varepsilon \boldsymbol{u})}{\partial t} + \nabla \cdot (\varepsilon \boldsymbol{u}\boldsymbol{u})$$
$$= -\frac{\varepsilon \nabla p}{\rho_f} + \frac{\mu_f}{\rho_f} \nabla \cdot \left(\varepsilon(\nabla \boldsymbol{u} + \nabla \boldsymbol{u}^T)\right) - \frac{\boldsymbol{f}_d}{\rho_f} + \varepsilon \boldsymbol{g} - \varepsilon \gamma (T_f - T_0) \boldsymbol{g}, \tag{16}$$

and

$$\frac{\partial \varepsilon}{\partial t} + \nabla \cdot (\varepsilon \boldsymbol{u}) = 0, \tag{17}$$

where $\boldsymbol{f}_d$, $\gamma$, $T_f$, and $T_0$ are the drag force acting on the fluid, thermal expansion coefficient, fluid temperature, and reference temperature, respectively. The drag force acting on the particle is given as:

$$\boldsymbol{f}_d = -\frac{\sum_{i=1}^{N_{grid}} \boldsymbol{F}_d}{V_{grid}}, \tag{18}$$

where $V_{grid}$, and $N_{grid}$ is the volume of the fluid grid, and the number of the solid particles located inside the fluid grid, respectively.

2.2. Heat transfer model

2.2.1. Energy equation

In the current study, the Eulerian framework is employed to model the heat transfer for solid phase in the DEM simulation. The governing equations of the heat transfer for fluid and solid phases are given as:

$$\frac{\partial(\varepsilon T_f)}{\partial t} + \nabla \cdot (\varepsilon T_f \boldsymbol{u}) = \frac{1}{\rho_f C_{p_f}} \nabla \cdot \left(\varepsilon k_f^{eff} \nabla T_f\right) - \frac{Q_{fs}}{\rho_f C_{p_f}} + \frac{Q_{fw}}{\rho_f C_{p_f}}, \qquad (19)$$

and

$$\begin{aligned}\frac{\partial((1-\varepsilon)T_s)}{\partial t} &+ \nabla \cdot ((1-\varepsilon)T_s \boldsymbol{v}^{CGM}) \\ &= \frac{1}{\rho_s C_{p_s}} \nabla \cdot \left((1-\varepsilon)k_s^{eff} \nabla T_s\right) + \frac{Q_{fs}}{\rho_s C_{p_s}} + \frac{Q_{sw}}{\rho_s C_{p_s}},\end{aligned} \qquad (20)$$

where $C_{p_f}$, $k_f^{eff}$, $T_s$, $\rho_s$, $C_{p_s}$, $k_s^{eff}$, $Q_{fs}$, $Q_{fw}$, and $Q_{sw}$ are the fluid specific heat, effective thermal conductivity, solid density, specific heat, effective thermal conductivity, heat transfer between fluid and particles, heat transfer from the wall to the fluid, and heat transfer from the wall to the solid, respectively. Our heat transfer model is newly developed by introducing the heat transfer model for Eulerian–Eulerian model [37–40] into the DEM–CFD method. Importantly, $k_s^{eff}$ is not influenced by the particle diameter and the stiffness of the solid particles. Accordingly, the heat transfer for the solid phase is influenced not by the contact state but by only the void fraction. This means that taking into consideration of the coarse grain ratio is not necessary in the heat transfer model for the solid phase, even when a coarse-grained DEM is employed.

2.2.2. Effective thermal conductivities

$k_f^{eff}$ in Eq. (19) and $k_s^{eff}$ in Eq. (20) can be modeled theoretically, and therefore, a typical theoretical model [41,42] is employed. The theoretical model has been delivered in a packed bed by considering heat conduction between solid particles as well as heat conduction through the fluid surrounding the solid particles. The thermal conductivity equation for solid-fluid mixture system is given as:

$$k_m^{eff} = \sqrt{1-\varepsilon}[\varphi k_s + (1-\varphi)\Gamma k_f] + (1-\sqrt{1-\varepsilon})k_f, \qquad (21)$$

$$\Gamma = \frac{2}{\left(1-\frac{B}{A}\right)}\left[\frac{(A-1)}{\left(1-\frac{B}{A}\right)^2}\frac{B}{A}\ln\left(\frac{A}{B}\right) - \frac{(B-1)}{\left(1-\frac{B}{A}\right)} - \frac{(B+1)}{2}\right], \qquad (22)$$

$$A = \frac{k_s}{k_f}, \qquad (23)$$

$$B = 1.25\left(\frac{1-\varepsilon}{\varepsilon}\right)^{\frac{10}{9}}, \qquad (24)$$

and

$$\varphi = 7.26 \times 10^{-3}, \qquad (25)$$

where $\varphi$, $k_f$, and $k_s$ are the parameter for the relative contact area of solid particles, the thermal conductivity of the fluid and the thermal conductivity of the solid, respectively. Adequacy of the theoretical model is experimentally proved in fluidized beds [43]. When the effective thermal conductivity, namely Eq. (21), is separately given to fluid and solid phases, the effective thermal conductivity of the fluid and solid phases should be respectively given as:

$$k_f^{eff} = \frac{1-\sqrt{1-\varepsilon}}{\varepsilon}k_f, \qquad (26)$$

and

$$k_s^{eff} = \frac{1}{\sqrt{1-\varepsilon}}[\varphi k_s + (1-\varphi)\Gamma k_f]. \qquad (27)$$

The above effective thermal conductivities for the fluid and solid phases are often employed in the Eulerian–Eulerian simulations, and the adequacy has been proved in the previous studies [38,39]. In addition, $\varphi$ is empirically given by Eq. (25) [38,39,42].

2.2.3. Solid-fluid heat transfer

In this study, the empirical model [44] is employed in the solid-fluid heat transfer. This model has been derived in packed and fluidized beds, and is available in a void fraction ranging from 0.35 to 1.0 as well as Reynolds number $\leq 10^5$. This empirical model is derived by incorporating several asymptotic conditions, and given by

$$\mathrm{Nu}_{fs} = (7 - 10\varepsilon + 5\varepsilon^2)\left(1 + 0.7\mathrm{Re}_p^{0.2}\mathrm{Pr}^{\frac{1}{3}}\right) \\ + (1.33 - 2.4\varepsilon + 1.2\varepsilon^2)\mathrm{Re}_p^{0.7}\mathrm{Pr}^{\frac{1}{3}}, \quad (28)$$

and

$$\mathrm{Pr} = \frac{\mu_f C_{p_f}}{k_f}, \quad (29)$$

where $\mathrm{Nu}_{fs}$, and Pr are the Nusselt number for heat transfer between solid and fluid phases, and Prandtl number, respectively. The empirical model has been widely employed in heat transfer simulations [38–40]. The heat transfer between solid and fluid phases is given by the particle surface area, the heat transfer coefficient, and the temperature difference between the particle fluids. Hence, $Q_{fs}$ in Eqs. (19) and (20) are given as:

$$Q_{fs} = \frac{\sum_{i=1}^{N_{grid}^O} \frac{\mathrm{Nu}_{fs} k_f}{d_p^O} S_p^O}{V_{grid}}(T_f - T_s) = l^2 \frac{\sum_{i=1}^{N_{grid}^{CGM}} \frac{\mathrm{Nu}_{fs} k_f}{d_p^{CGM}} S_p^{CGM}}{V_{grid}}(T_f - T_s), \quad (30)$$

where $S_p$ is the surface area of the particle.

2.2.4. Solid-wall and fluid-wall heat transfer

In this study, the heat fluxes for the wall-to-fluid phase and the wall-to-solid phase are modeled as the heat sources in the energy equations, namely $Q_{fw}$ in Eq. (19) and $Q_{sw}$ in Eq. (20). By applying the divergence theorem, the heat fluxes can be converted to the heat sources give as:

$$Q_{fw} = -\nabla \cdot \boldsymbol{q}_{fw}, \quad (31)$$

and

$$Q_{sw} = -\nabla \cdot \boldsymbol{q}_{sw}, \quad (32)$$

where $\boldsymbol{q}_{fw}$, and $\boldsymbol{q}_{sw}$ are the heat flux for fluid-wall, and heat flux for solid-wall, respectively. The heat sources are given to the grids including the wall surface. Based on

Newton's cooling law, the heat flux for fluid-wall and solid-wall are given as:

$$\boldsymbol{q}_{fw} = \varepsilon h_w(T_f - T_{ext})\boldsymbol{n}, \tag{33}$$

and

$$\boldsymbol{q}_{sw} = (1-\varepsilon)h_w(T_s - T_{ext})\boldsymbol{n}, \tag{34}$$

where $h_w$, $T_{ext}$, and $\boldsymbol{n}$ are the wall heat transfer coefficient, the temperature of the external environment, and unit normal vector at wall surface, respectively.

## 3. Validation test

### 3.1. Simulation conditions

To prove the adequacy of the proposed heat transfer model, a simulation for the cooling of solid particles in a fluidized bed was performed under conditions comparable to those of the previous experiment [21,45]. Indeed, experimental data for validation tests in granular dynamics involving heat transfer were extremely limited. Therefore, this system [21,45] was selected because the experimental conditions [45] corresponding to the simulation [21] were available. Fig. 1 shows the calculation domain and the initial location of the particles. The size of the calculation domain was 80 mm × 250 mm × 15 mm. The particle diameter was 1.0 mm, and the total weight of the particles was 75 g. Table 1 presents the physical properties used in the validation test. As shown in Table 2, two cases of the simulation were performed with different inlet velocities. The gas was injected from the bottom side at superficial velocities of 1.20 m/s in Case 1-1 and 1.71 m/s in Case 1-2. The solid particles whose initial temperature was 90 °C were randomly located in the fluidized bed, and then cooled by injected gas whose temperature was 20 °C. The grid size and the time step were set to 2.5 mm and $1.0\times10^{-5}$ s, respectively. In the same way as the previous study [21], the reasonable heat transfer coefficients were carefully chosen to reproduce the temperature distribution corresponding to the experimental results [45], namely 60 W/m$^2$ K. The same heat transfer coefficients were used in Case 1-1 and Case 1-2. At the inlet, a fixed temperature boundary condition was applied for the gas, and an adiabatic boundary condition was applied for the solid particles. Thus, the validation tests were performed under the fair conditions of previous studies [21,45].

3.2. Results and discussion

First, the validation test for Case 1-1 is presented. Fig. 2 shows the spatial particle location, and solid temperature distribution at 1.0 s. The particle behavior and temperature distribution were qualitatively consistent with the experimental results [45]. As shown in Fig. 2 (a), the particles were fluidized, and bubbles were observed in the bulk region. Fig. 2 (b) indicates that the temperature of the solid phase was lower at the bottom side of the fluidized bed as well as at the center of the bulk region due to the upward movement of particles cooled by the injected gas Subsequently, quantitative comparisons were made in Case 1-1. Fig. 3 illustrates the transient changes of the average temperature of the solid phase obtained from the simulation and the experiment [45]. The solid temperature gradually decreased due to the injection of low-temperature gas, and besides, the solid temperatures well agreed between the computation and the experiment [45]. The vertical temperature distribution of the solid phase was compared between the simulation and the experiment [45]. Herein, y ≥ 10 mm was selected as the comparison region because this region was regarded to be reliable from the previous study [21]. As shown in Fig. 4, the calculated temperature quantitatively agreed with the experimental results [45]. Thus, the adequacy of the proposed heat transfer model was proved. This implies that the new heat transfer model could reproduce the temperature of the solid phase with the simple setup, namely, giving the void fraction in the heat transfer model.

Next, the validation test results for Case 1-2 are described. The results of Case 1-2 were similar to those of Case 1-1. Fig. 5 shows the spatial particle location, and the solid temperature distribution at 1.0 s. As shown in Fig. 5 (a), the particles were fluidized and moved more vigorously compared to those in Case 1-1. As shown in Fig. 5 (b), a lower-temperature part was observed at the center of the bulk, but the temperature difference between the higher and lower parts was smaller than that in Case 1-1. Thus, the particle behavior and the temperature distribution qualitatively agreed between the calculation and the experiment [45]. A quantitative comparison of the temperature distribution of the solid phase was conducted between the simulation and the experiment [45]. Fig. 6 shows that transient changes of the average solid temperature agreed well between the simulation and the experiment [45]. As shown in Fig. 7, the temperature distribution along the y-axis (y ≥ 10 mm) agreed well between the computation and the experiment [45]. Thus, the adequacy of the proposed heat transfer model was shown. Again, the new heat transfer model made it possible to reproduce the actual temperature by only giving the void fraction.

Consequently, the adequacy of the proposed heat transfer model was shown through the validation tests. This means that heat transfer for the solid phase could be modeled using the Eulerian framework in the DEM–CFD method.

4. Sensitivity analysis for stiffness
4.1. Simulation conditions

To investigate the applicability of the soft spring model to the new heat transfer model, a sensitivity analysis was performed, where the spring constant value was used as the parameter. In the existing heat transfer model for the DEM, the contact area and duration are estimated since these parameters significantly affect the temperature distribution. This implies that the spring constant value was critical. In the DEM, soft spring is usually employed. This implies that the spring constant value is given flexibly. Accordingly, the heat transfer model which is not influenced by the spring constant is essential. Under the same conditions as those in Case 1-1, three types of simulations were performed in the sensitivity analysis. Table 3 summarizes the calculation conditions. In the sensitivity analysis, spring constants of $1.0 \times 10^2$ N/m, $1.0 \times 10^3$ N/m, and $1.0 \times 10^4$ N/m were used in Case 2-1, Case 2-2, and Case 2-3, respectively. The other physical properties were the same as those in Case 1.

4.2. Results and discussion

The sensitivity analysis results were examined at 1.0 s. First, the solid particle locations were compared. As shown in Fig. 8, the macroscopic characteristics of the gas-solid flow such as the height of the particle bed and the size of the bubbles were in qualitative agreement in Cases 2-1 to 2-3, where the different values for the spring constant were applied. Next, the distributions of void fraction and fluid velocity were compared. As illustrated in Fig. 9, the void fractions and fluid velocities corresponded well in Cases 2-1 to 2-3. Fig. 10 shows the temperature distribution of the fluid. The fluid temperatures were quite similar in Cases 2-1 to 2-3. The temperature distributions of the solid particles were compared as shown in Fig. 11. The temperature distributions were in good agreement in Case 2-1 to Case 2-3 despite the different spring constant values. These results indicate that the spring constant value rarely influenced the macroscopic behavior of solid particles such as the particle location, void fraction, and fluid velocity as shown in the past validation tests for the soft spring model. Moreover, the spring constant value was shown not to affect the temperature distribution because the new heat transfer model was developed based on the solid volume fraction. Next, a quantitative evaluation of the solid temperature distribution was examined. Fig. 12 illustrates the transient change of the average temperature of the solid phase. The average temperature of the solid phase agreed well in Cases 2-1 to 2-3. The temperature distribution of the solid particles along the vertical direction is shown in Fig. 13. The solid temperature distributions along the vertical direction quantitatively agreed in Cases 2-1 to 2-3. Thus, the temperature distribution obtained from the new heat transfer model was shown to be hardly influenced by the spring constant values unlike the conventional heat transfer model.

5. Compatibility of the coarse-grained DEM

5.1. Simulation conditions

Finally, to examine the compatibility of the coarse-grained DEM with the proposed heat transfer model, two cases of the numerical simulation were performed in two types of fixed beds with different void fractions. To assess the resolution of the solid phase, this compatibility analysis investigates the effect of the coarse-grain ratio on the temperature distribution. The computational domain and initial particle state are shown in Fig. 14. The physical properties were the same as those in the above computations, namely Table 1. The computational domain was filled with the fixed solid particles. The particles were aligned on a primitive cubic lattice, where the particles were in point contact, and hence, the void fraction was estimated to be 0.476 in Case 3-1. In Case 3-2, the particles were aligned on a body-centered cubic lattice, where the particles were not in contact, and hence, the void fraction was estimated to be 0.869. In these calculations, the solid particles were cooled by the gas injected at superficial velocity of 1.0 m/s. Table 4 shows the calculation conditions. The validation tests were performed to confirm the coarse-grained DEM with the new heat transfer model could reproduce the macroscopic characteristics of the temperature distribution. Case 3-1-1 was the original particle system, where the particle diameter was 1.0 mm, and the number of particles was 320,000. In Case 3-1-2 and Case 3-1-3, the coarse-grained DEM was employed, where the coarse grain ratio was 2.0 and 4.0. By employing the coarse-grained DEM, the number of the computational particles could be reduced drastically, and hence, the number of the computational particles was 40,000 and 5,000 in Case 3-1-2 and Case 3-1-3. Case 3-1-4 was performed by giving the particle diameter to be 4.0 mm without the coarse-grained DEM. The effectiveness of the coarse-grained DEM was examined by comparing Case 3-1-4 with Case 3-1-3, where the computational particle sizes were equivalent in these cases. Case 3-2 was performed in the same way as Case 3-1. Case 3-2-1 was original particle system. In Case 3-2-2 and Case 3-2-3, the coarse-grained DEM was employed. The effectiveness of the coarse-grained DEM was examined by comparing Case 3-2-4 with Case 3-2-3. In Cases 3-1 and 3-2, the grid size was set to 6.0 mm, and the time step was set to $1.0 \times 10^{-4}$ s.

5.2. Results and discussion

First, the validation test for Case 3-1 was addressed. Fig. 15 shows the solid particle temperature distribution. The solid phase temperature became lower at the left side due to the injected cold gas whose temperature was 20 °C. From the temperature distribution obtained, it was confirmed that the coarse-grained DEM incorporating the new heat transfer model (Cases 3-1-2 and 3-1-3) could reproduce the temperature distribution of the original particle systems (Case 3-1-1). On the other hand, the temperature of the simply large-sized particle system (Case 3-1-4) was entirely different from that of the coarse-grained particle system (Case 3-1-3), despite their computational particle sizes being equivalent. Thus, these results showed that the coarse-grained DEM successfully reproduced the temperature distribution of the solid phase. However, when simply large-sized particles were used without the coarse-grained DEM, the simulation results became significantly different from the expected results, namely, Case 3-1-1.

Accordingly, these results imply that the new heat transfer model was shown to have high compatibility with the coarse-grained DEM. Subsequently, quantitative comparisons were conducted between the original particle system and the coarse-grained DEM in Cases 3-1-1 to 3-1-3. As shown in Fig. 16, the solid temperature distribution along the x-axis direction agreed well between the original particle system (Case 3-1-1) and the coarse-grained DEM (Cases 3-1-2 and 3-1-3). Importantly, the error due to the new heat transfer model was proved to be extremely small, even when the heat transfer model was incorporated into the coarse-grained DEM. Thus, the proposed heat transfer model precisely reproduced the temperature distribution of the solid phase, even when using the coarse-grained DEM. In particular, the coarse-grained DEM incorporating the new heat transfer model could reproduce the original particle systems simply by providing the void fraction, without requiring a complex procedure. Therefore, the new heat transfer model completely resolves critical problems in the existing DEM simulations involving heat transfer.

Next, the validation test results for Case 3-2 were examined. The results of Case 3-2 were similar to those of Case 3-1. Fig. 17 shows the temperature distributions of the solid phase. The solid temperature distributions were in good agreement across Cases 3-2-1 to 3-2-3, whereas the temperature obtained from Case 3-2-4 was different from the other cases. These results indicate that the temperature distribution in the original particle system (Case 3-2-1) could be qualitatively reproduced using the coarse-grained particle system (Cases 3-2-2 and 3-2-3). In contrast, simply using large particle system (Case 3-2-4) fails to replicate the original system. Subsequently, quantitative evaluations were performed in Cases 3-2-1 to 3-2-3. Fig. 18 shows the temperature distribution of solid phase, and it was in good agreement in Cases 3-2-1 to 3-2-3. Again, the error due to incorporating the heat transfer model into the coarse-grained DEM was shown to be extremely small. Thus, the temperature distribution in the original system could be quantitatively reproduced by the coarse-grained DEM incorporating the new heat transfer model. The success of this approach indicates that the new heat transfer model, which accounts for the solid volume fraction, resolves critical limitations in traditional DEM simulations involving heat transfer.

In summary, the coarse-grained DEM with the new heat transfer model successfully reproduced the solid phase temperature distribution. Notably, the integration of the new heat transfer model into the coarse-grained DEM allowed the reproduction of original particle system by a simplified approach, relying solely on the void fraction without contact state modeling. This innovative heat transfer model resolves longstanding challenges in conventional DEM simulations involving heat transfer, offering a robust and efficient solution for simulating large-scale powder systems involving heat transfer.

6. Conclusions

In the existing heat transfer model for the DEM, a Lagrangian approach was employed in the solid phase, where complex modeling was employed to correspond to the soft spring model. However, the existing heat transfer model had critical problems: the calculation results were influenced by the stiffness of the computational particles and the coarse grain ratio. To resolve these essential problems, we developed a new heat transfer model for the DEM, where an Eulerian framework was incorporated into the DEM simulation. The new heat transfer model was uniquely constructed in the DEM–CFD method by considering the volume fraction, and hence, could be implemented without complex modeling. Importantly, the new heat transfer model was theoretically unaffected by the spring constant value for the solid particles and the coarse grain ratio. To show the adequacy of the proposed heat transfer model, we conducted several validation tests and numerical simulations. The results demonstrated that the new heat transfer model could successfully reproduce the temperature distribution in gas-solid flows with both high accuracy and universality. Besides, the new heat transfer model was shown to be independent of the spring constant values as well as coarse grain ratios. Thus, the adequacy of the new heat transfer model was proved by agreement between the computational and the experimental results, and besides, the calculated temperature of the solid phase was shown not to be influenced by the spring constant value. In addition, the new heat transfer model was shown to have high compatibility with the coarse-grained DEM. Consequently, the essential problems of the heat transfer model for the DEM simulation were completely resolved by our proposed model.

Acknowledgements

The authors acknowledge the financial support from the Japan Society for the Promotion of Science KAKENHI (Grant Nos. 21H04870 and 24K22289).

Declaration of generative AI and AI-assisted technologies in the writing process

During the preparation of this work the author(s) used ChatGPT in order to improve language and readability. After using this tool/service, the authors reviewed and edited the content as needed and take full responsibility for the content of the publication.

Declaration of competing interest

The authors declare that they have no known competing financial interests or personal relationships that could have appeared to influence the work reported in this paper.

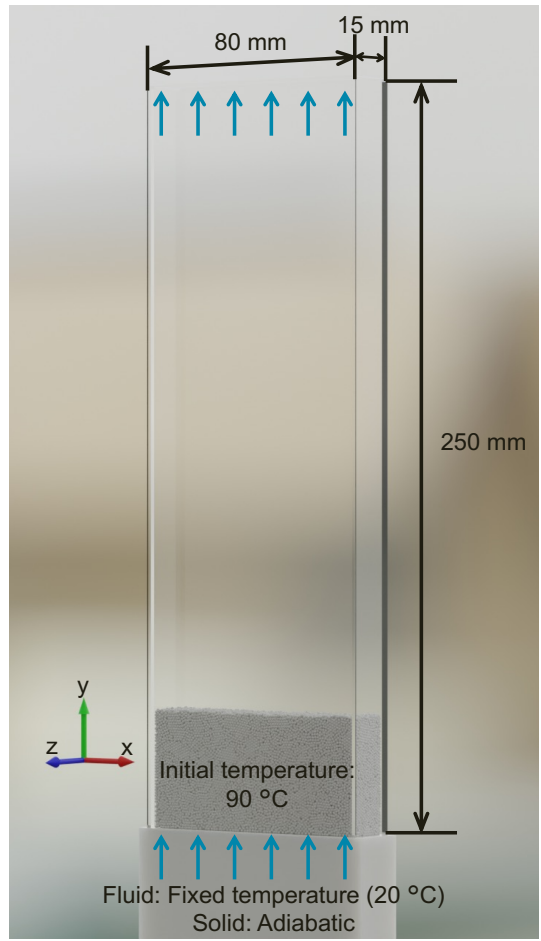

Fig. 1 Simulation setup of fluidized bed.

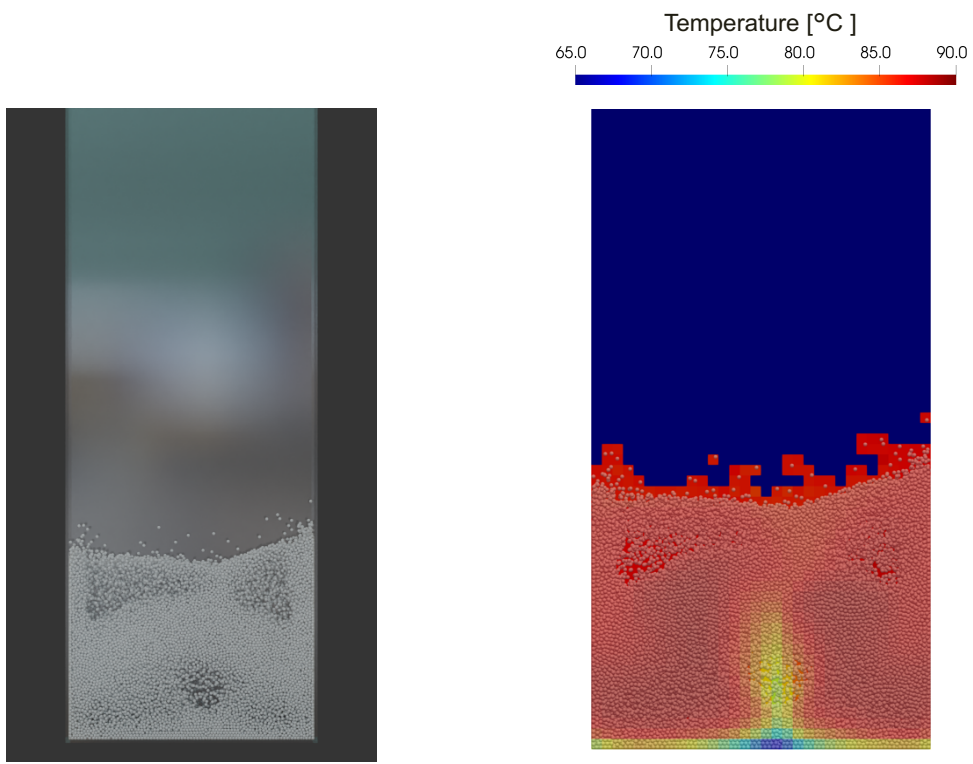

(a) Particle location  (b) Solid temperature

Fig. 2 Snapshots at validation test result for Case 1-1 (1.0 s).

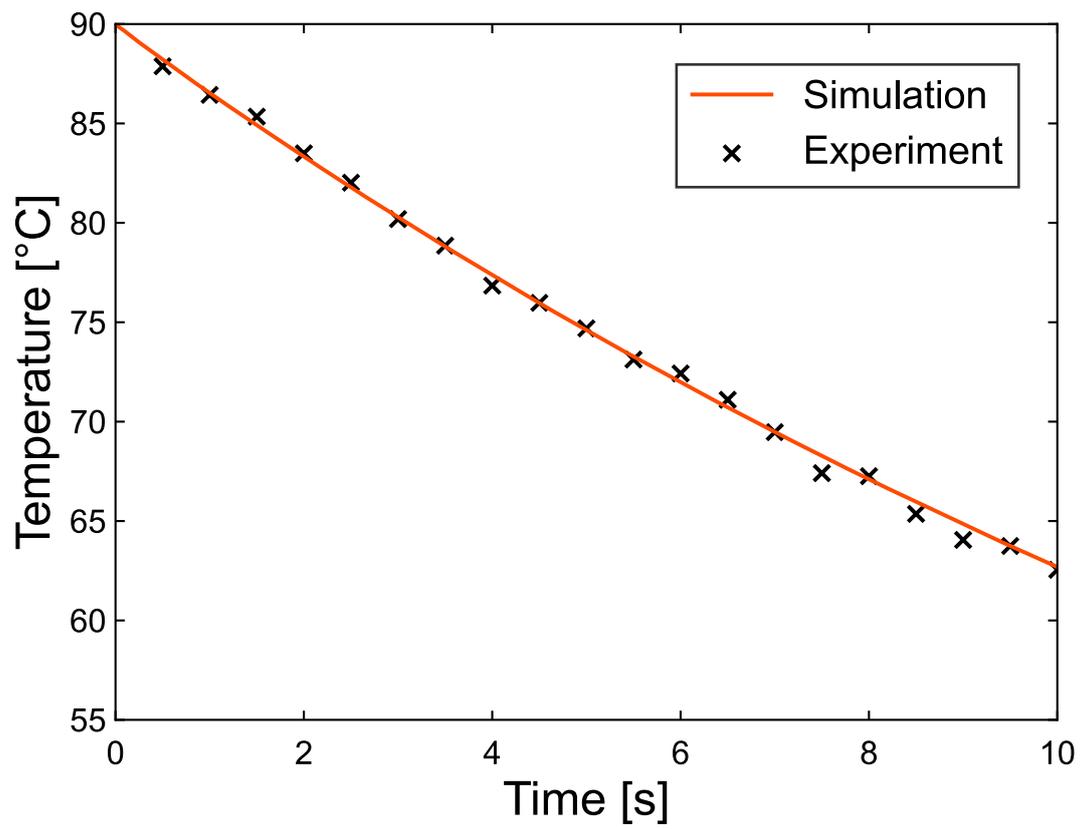

Fig. 3 Comparison of average temperature for solid phase region between the simulation and the experiment in Case 1-1.

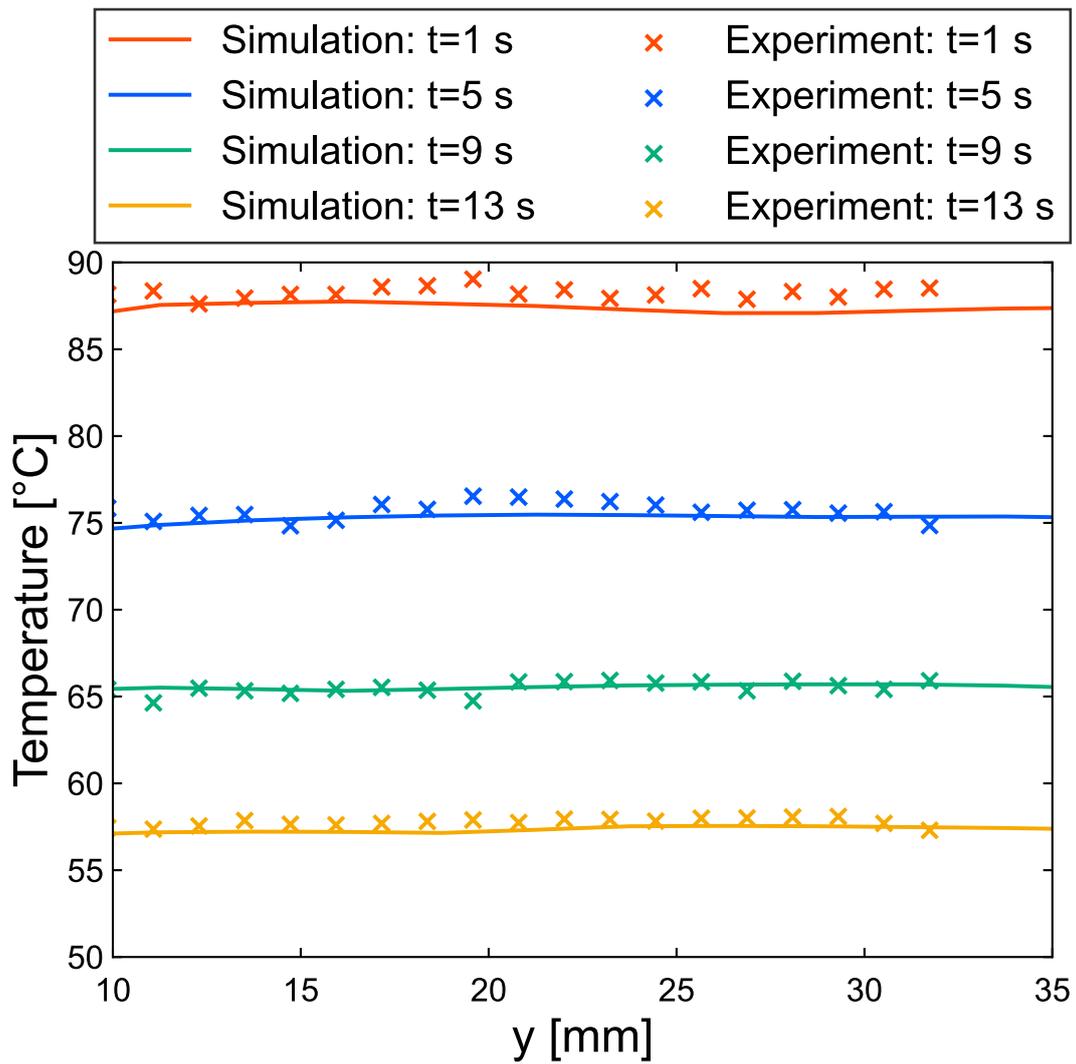

Fig. 4 Comparison of solid temperature change along y-axis in the simulation and the experiment for Case 1-1.

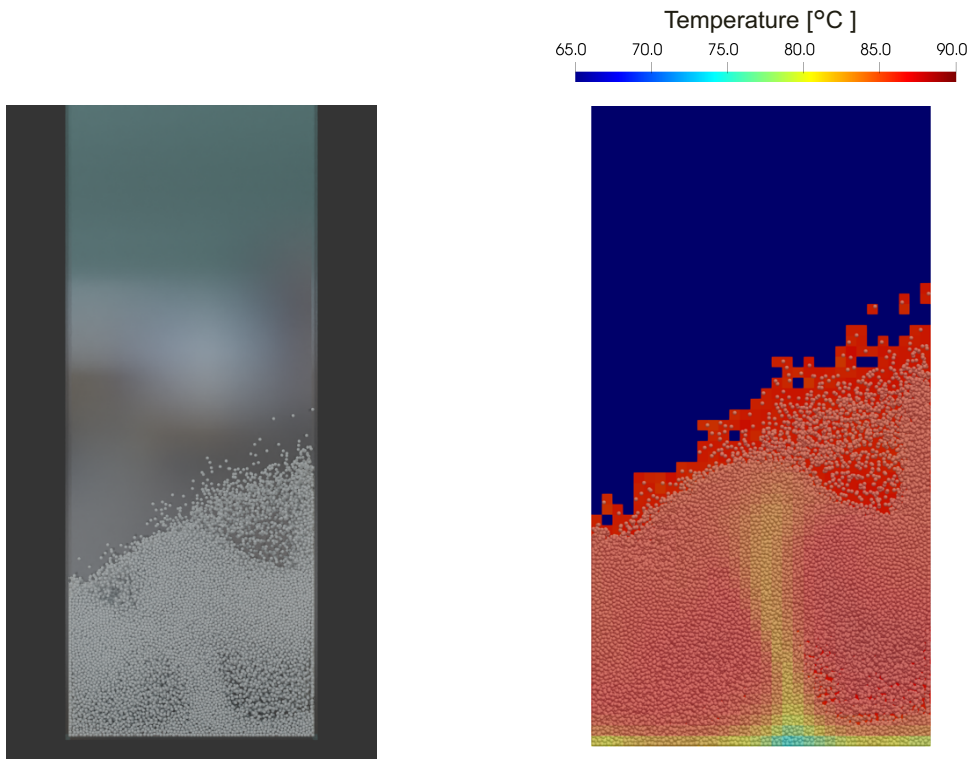

(a) Particle location    (b) Solid temperature

Fig. 5 Snapshots at validation test result for Case 1-2 (1.0 s).

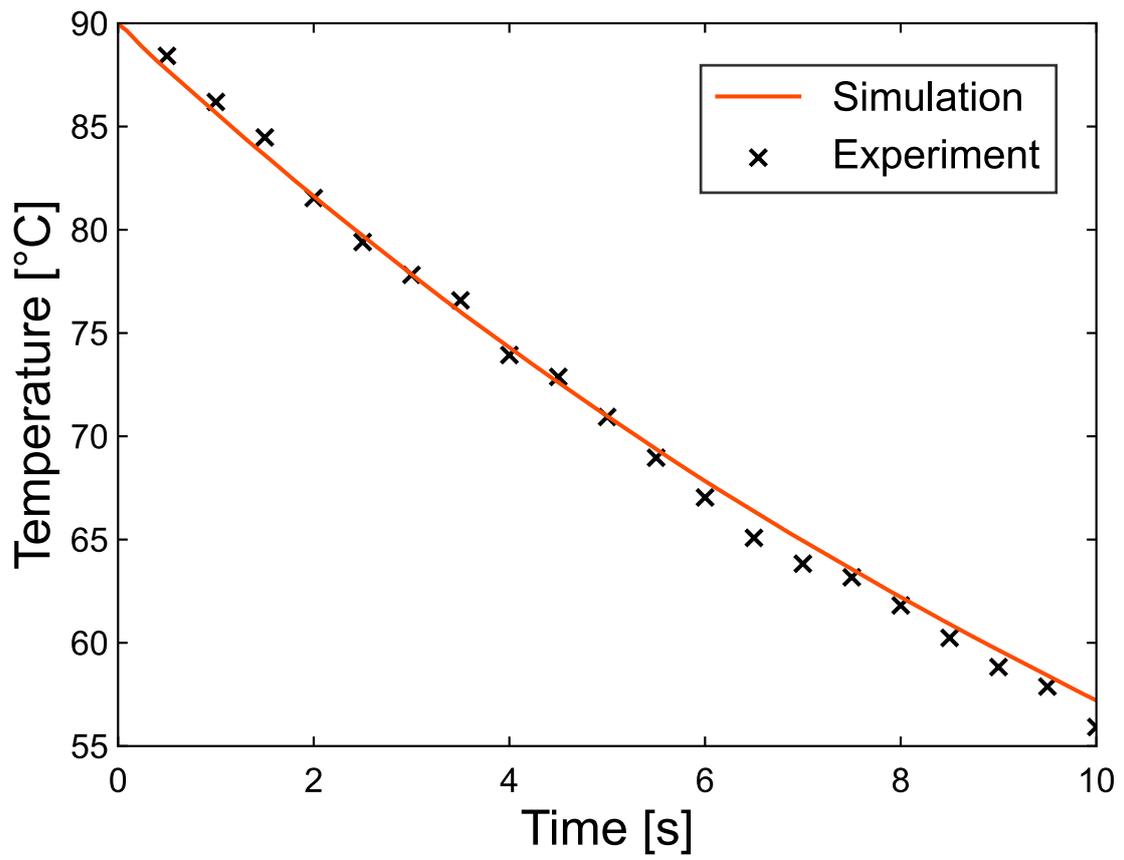

Fig. 6 Comparison of average temperature of the solid phase region between the simulation and the experiment in Case 1-2.

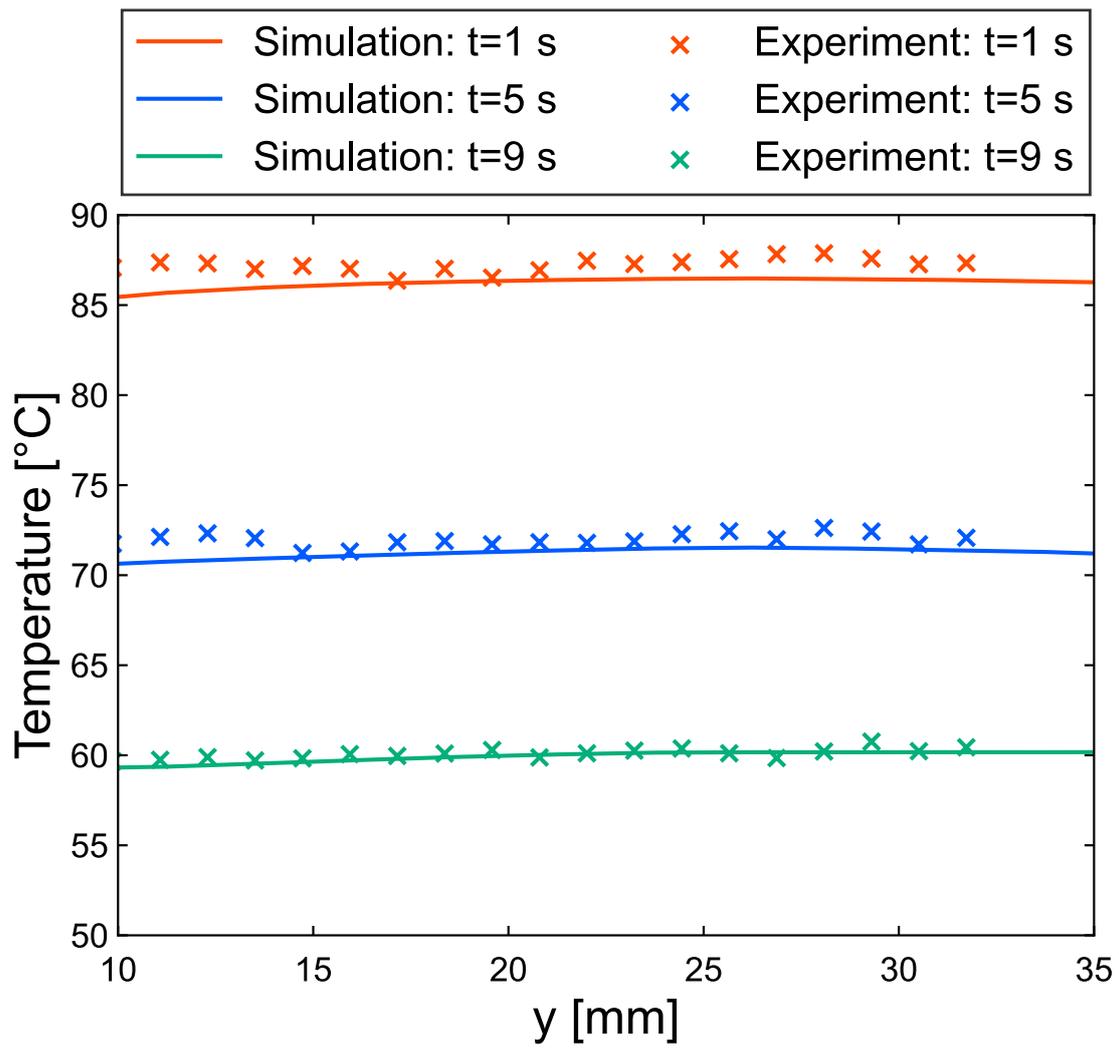

Fig. 7 Comparison of change of the solid temperature along y-axis in the simulation and the experiment for Case 1-2.

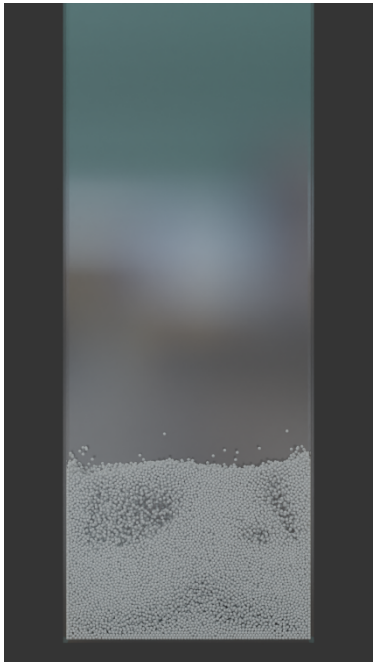 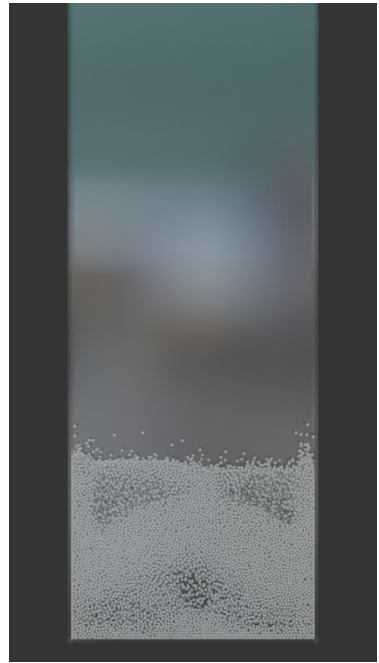 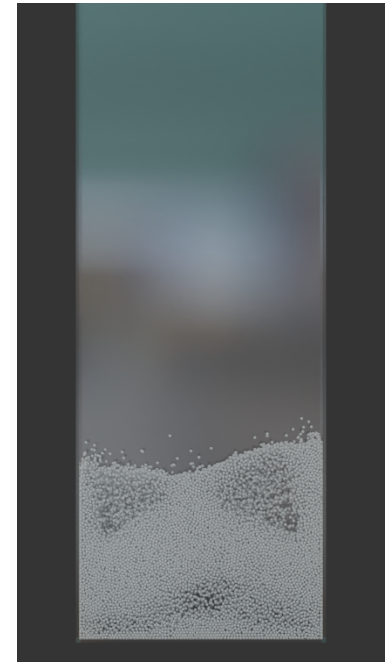

(a) Case 2-1　　　　　　　　　　　　(b) Case 2-2　　　　　　　　　　　　(c) Case 2-3

Fig. 8　Snapshots for particle location in Case 2 (1.0 s).

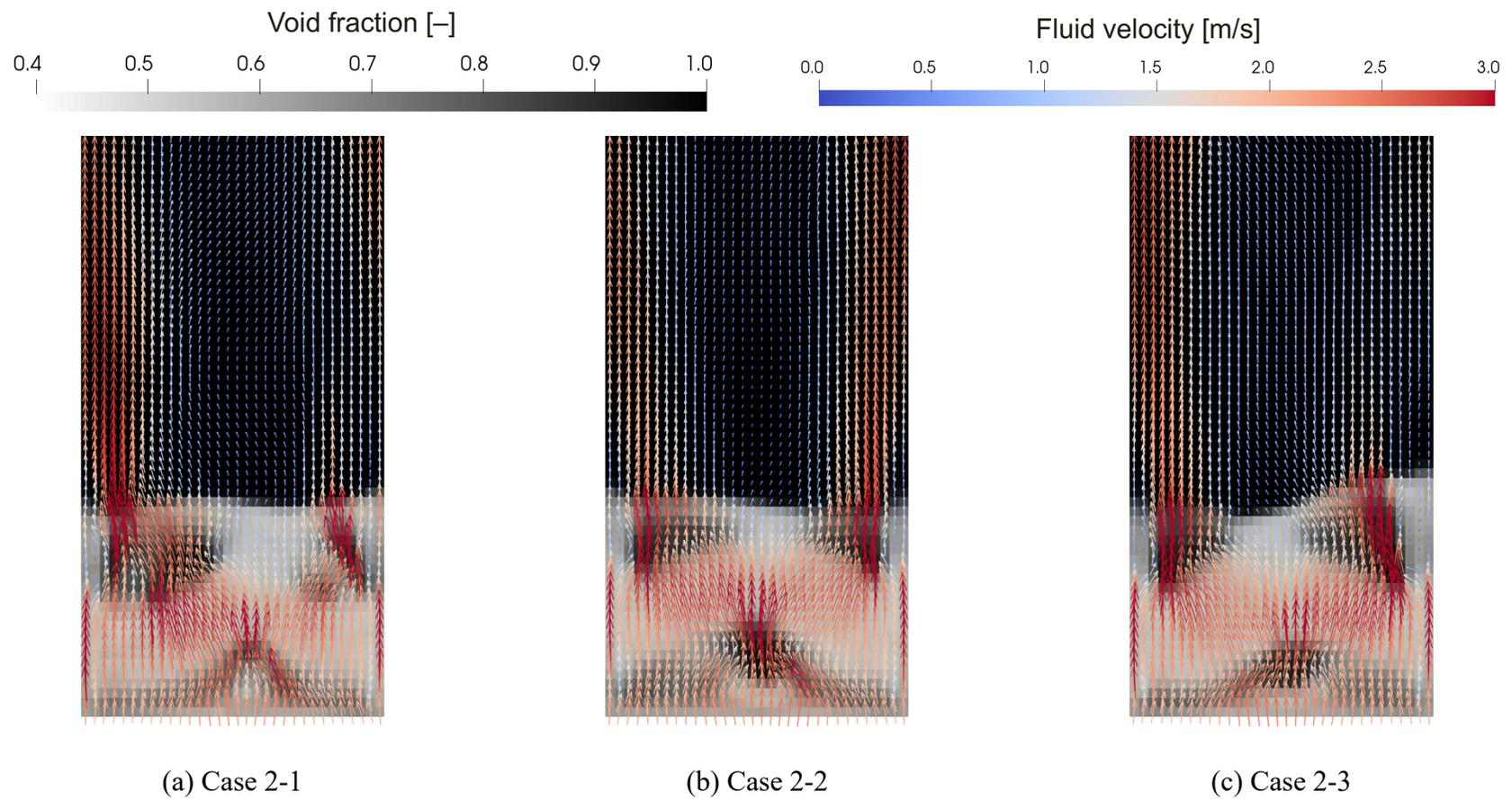

(a) Case 2-1  (b) Case 2-2  (c) Case 2-3

Fig. 9 Snapshots for void fraction and fluid velocity distribution in Case 2 (1.0 s).

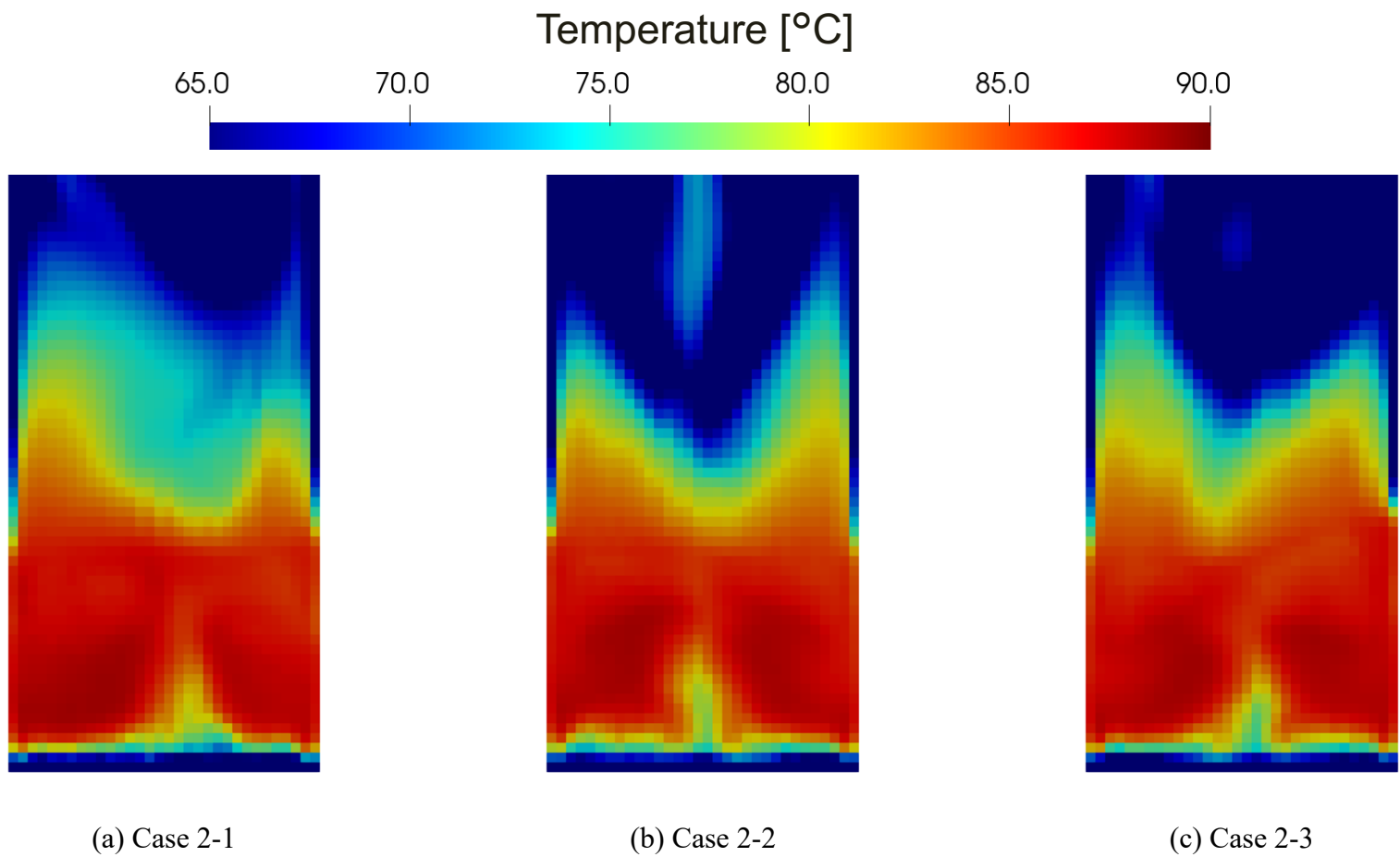

(a) Case 2-1  (b) Case 2-2  (c) Case 2-3

Fig. 10 Snapshots of fluid temperature in Case 2 (1.0 s).

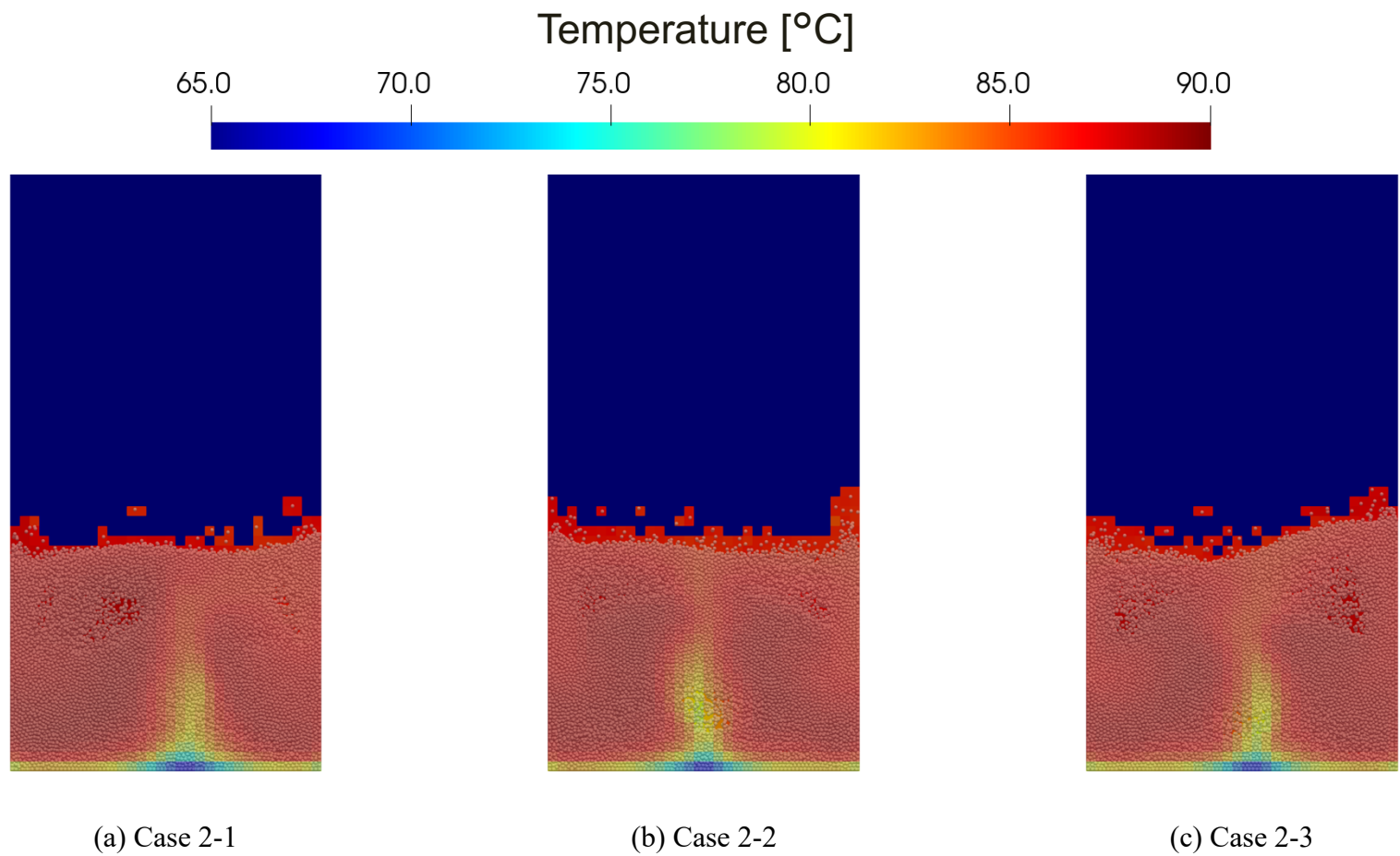

(a) Case 2-1  (b) Case 2-2  (c) Case 2-3

Fig. 11 Snapshots of solid temperature in Case 2 (1.0 s).

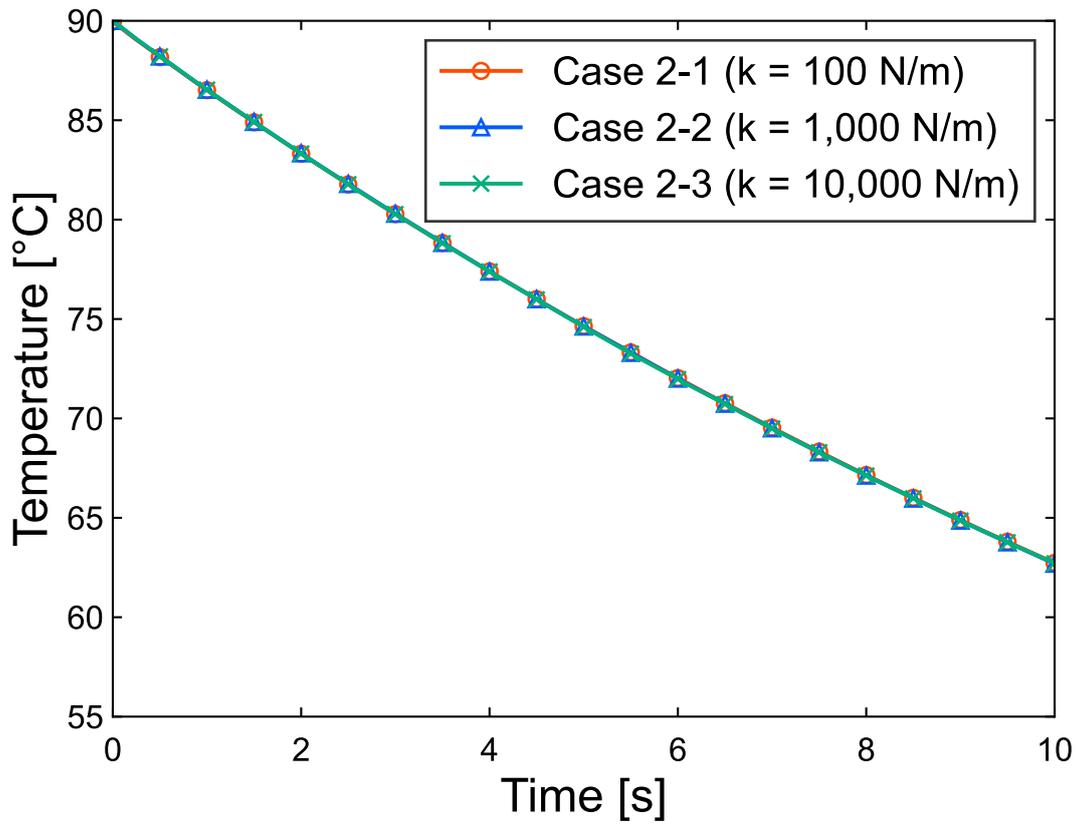

Fig. 12 Change of average temperature of the solid phase in Case 2.

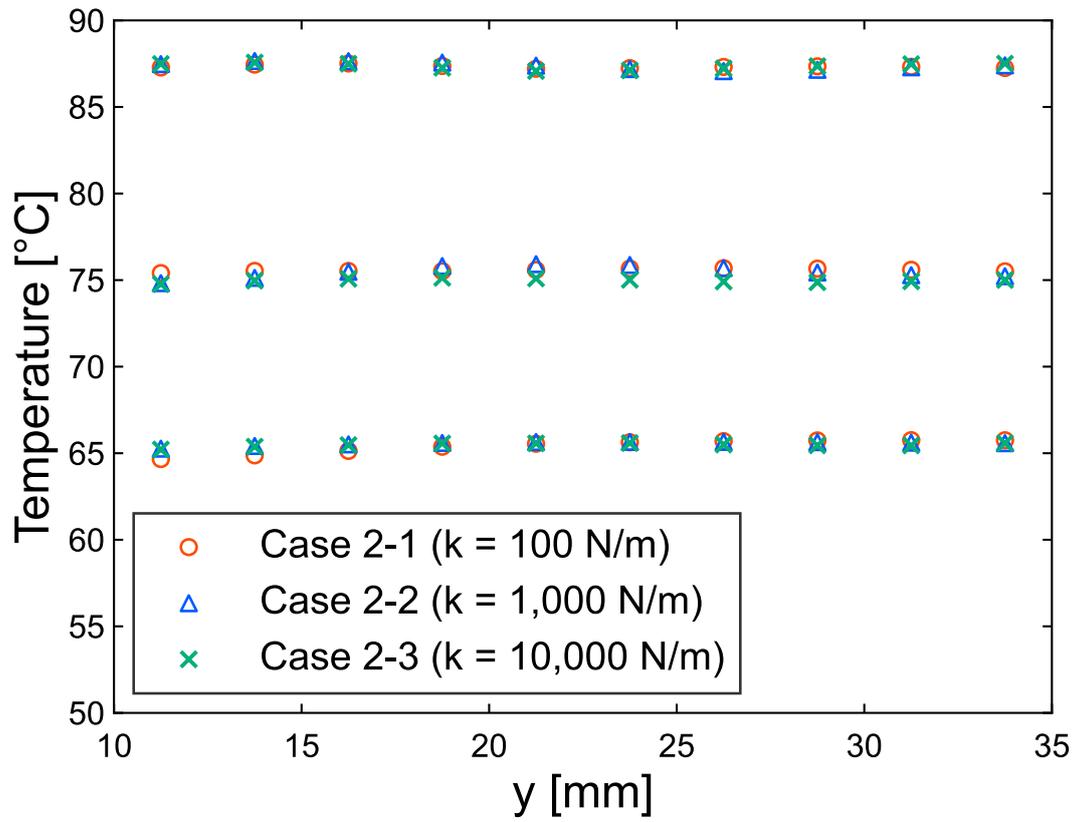

Fig. 13 Change of solid temperature distribution along y-axis in Case 2.

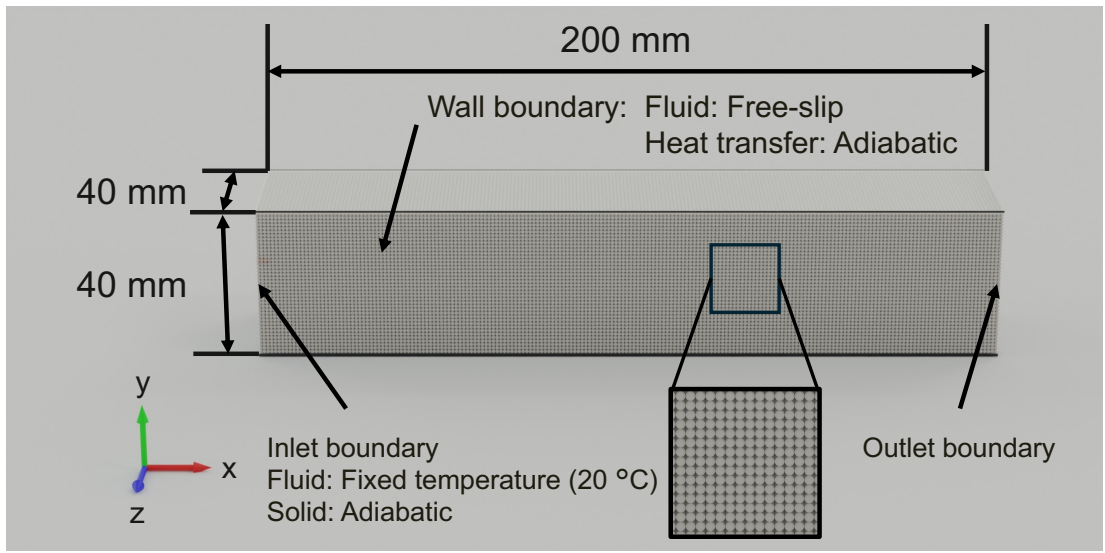

(a) Case 3-1

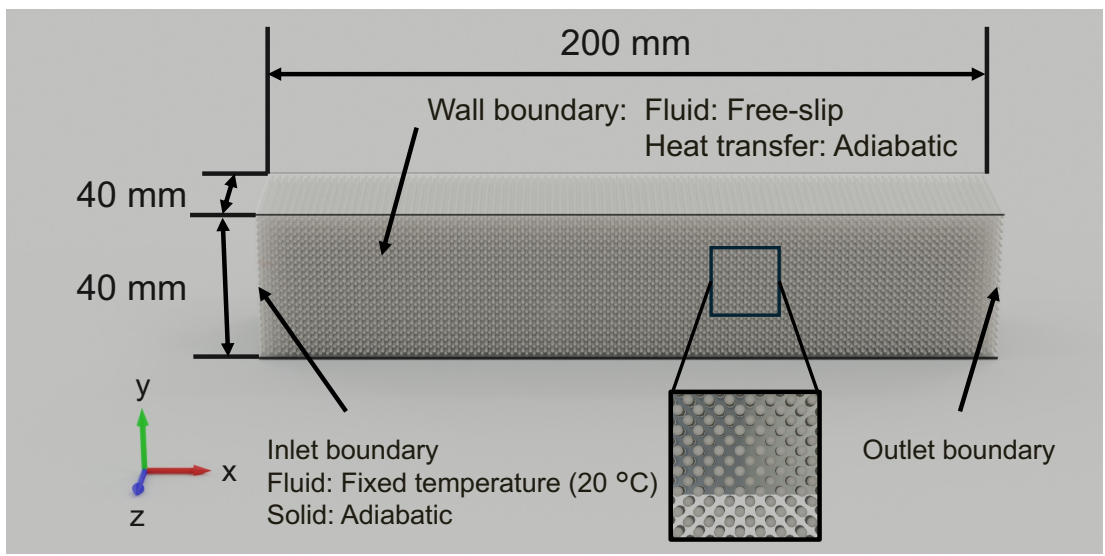

(b) Case 3-2

Fig. 14 Simulation setup of fixed bed.

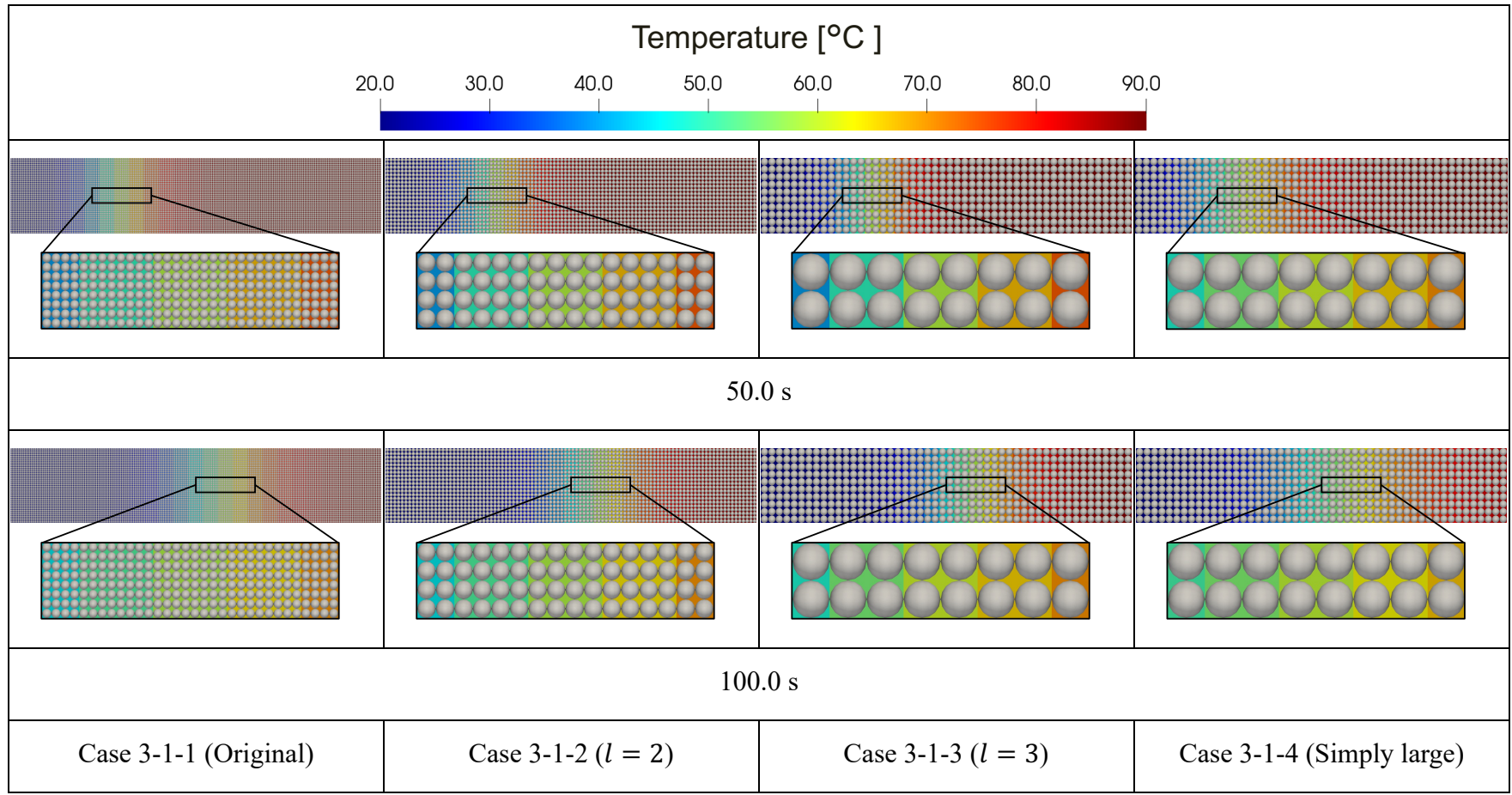

Fig. 15 Temperature distribution of solid phase for Case 3-1.

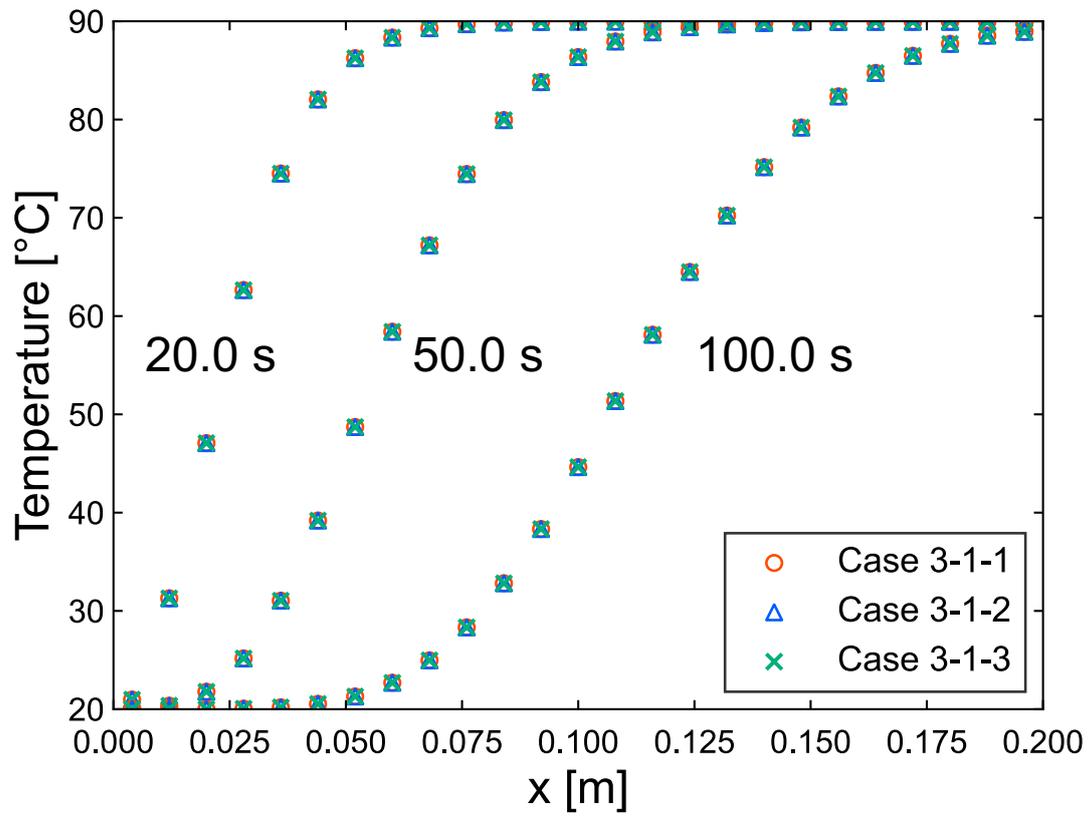

Fig. 16 Solid temperature distribution along x-axis for Case 3-1.

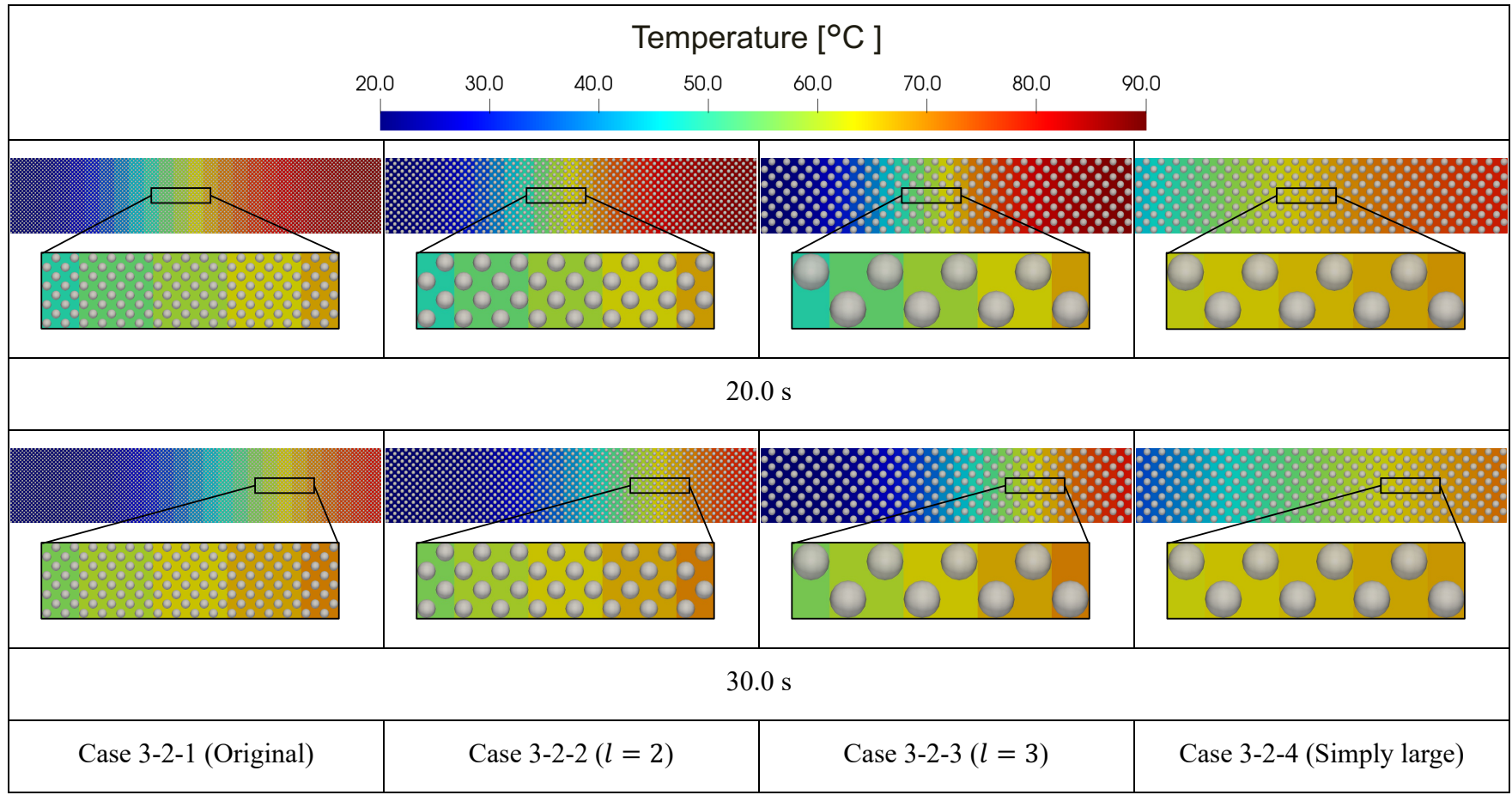

Fig. 17 Temperature distribution of solid phase for Case 3-2.

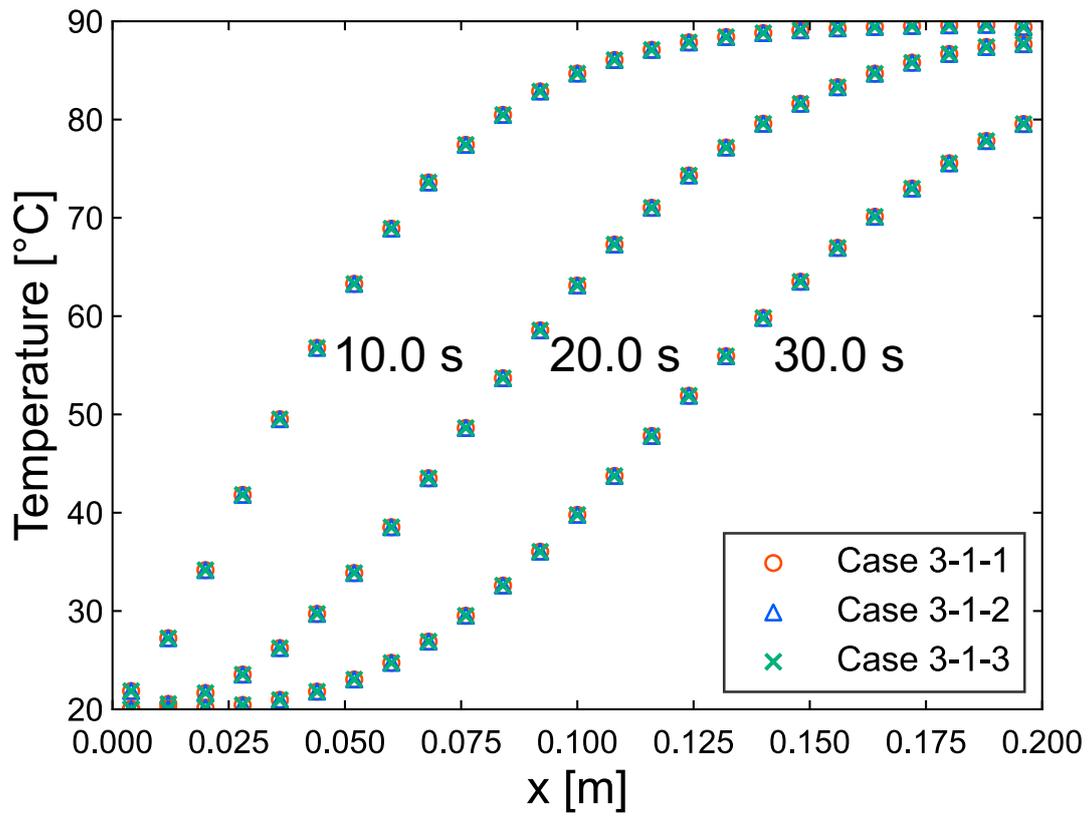

Fig. 18 Solid temperature distribution along x-axis for Case 3-2.

Table 1 Physical properties.

| | |
|---|---|
| *Solid property* | |
| Spring constant [N/m] | $1.0\times10^3$ |
| Restitution coefficient [–] | 0.97 |
| Friction coefficient [–] | 0.3 |
| Density [kg/m$^3$] | 2500 |
| Specific heat capacity [J/kg K] | 840 |
| Thermal conductivity [W/m K] | 1.4 |
| *Gas property* | |
| Density [kg/m$^3$] | 1.2 |
| Viscosity [Pa s] | $2.0\times10^{-5}$ |
| Specific heat capacity [J/kg K] | 1010 |
| Thermal conductivity [W/m K] | 0.025 |
| Coefficient of thermal expansion [1/K] | 0.003 |
| Heat transfer coefficient [W/m$^2$ K] | 60 |

Table 2 Simulation cases for validation test.

|  | Case 1-1 | Case 1-2 |
|---|---|---|
| Particle size [mm] | 1.0 | |
| Number of particles [–] | 57,296 | |
| Grid size [mm] | 2.5 | |
| Time step [s] | $1.0\times10^{-5}$ | |
| Superficial velocity [m/s] | 1.20 | 1.71 |

Table 3 Simulation cases for sensitivity analysis of the spring constant value.

|  | Case 2-1 | Case 2-2 | Case 2-3 |
|---|---|---|---|
| Particle size [mm] | | 1.0 | |
| Number of particles [–] | | 57,296 | |
| Grid size [mm] | | 2.5 | |
| Time step [s] | | $2.5\times10^{-6}$ | |
| Superficial velocity [m/s] | | 1.20 | |
| Spring constant [N/m] | $1.0\times10^{2}$ | $1.0\times10^{3}$ | $1.0\times10^{4}$ |

Table 4 Simulation cases for validation test to confirm the compatibility of the coarse-grained DEM.

| | Case 3-1-1 | Case 3-1-2 | Case 3-1-3 | Case 3-1-4 | Case 3-2-1 | Case 3-2-2 | Case 3-2-3 | Case 3-2-4 |
|---|---|---|---|---|---|---|---|---|
| Coarse grain ratio [–] | 1.0 | 2.0 | 4.0 | 1.0 | 1.0 | 2.0 | 4.0 | 1.0 |
| Original particle size [mm] | 1.0 | 1.0 | 1.0 | 4.0 | 1.0 | 1.0 | 1.0 | 4.0 |
| Computational particle size [mm] | 1.0 | 2.0 | 4.0 | 4.0 | 1.0 | 2.0 | 4.0 | 4.0 |
| Number of computational particles [–] | 320,000 | 40,000 | 5,000 | 5,000 | 80,000 | 10,000 | 1,250 | 1,250 |
| Grid size [mm] | 8.0 | | | | | | | |
| Time step [s] | $1.0\times10^{-5}$ | | | | | | | |